\documentclass[aps,preprint,nofootinbib]{revtex4}
\usepackage{graphics}
\usepackage{slashed}
\usepackage{amssymb}
\usepackage{lscape}
\usepackage{amsmath}
\usepackage{rotating}
\usepackage{graphicx}
\graphicspath{ {images/} }
\begin{document}
\title{Towards unified spinor fields: confinement of gravitons on a dS background}
\author{$^{2}$ Marcos R. A. Arcod\'{\i}a\footnote{E-mail address: marcodia@mdp.edu.ar},  $^{1,2}$ Mauricio Bellini
\footnote{E-mail address: mbellini@mdp.edu.ar} }
\address{$^1$ Departamento de F\'isica, Facultad de Ciencias Exactas y
Naturales, Universidad Nacional de Mar del Plata, Funes 3350, C.P.
7600, Mar del Plata, Argentina.\\
$^2$ Instituto de Investigaciones F\'{\i}sicas de Mar del Plata (IFIMAR), \\
Consejo Nacional de Investigaciones Cient\'ificas y T\'ecnicas
(CONICET), Mar del Plata, Argentina.}
\begin{small}
\begin{abstract}
We propose an unified theory for spinor fields on extended Weyl manifolds taking into account self-interactions to obtain the Relativistic dynamics on a general curved Riemannian background as continuation of the Relativistic Quantum Geometry program, recently introduced. We focuss our attention separately on both, massless and  matter fields. We study an example of confined gravitons on a de Sitter (dS) background at Planckian scales.
\end{abstract}
\maketitle

\section{INTRODUCTION}

One of the central problems in contemporary theoretical physics is the unification of quantum field theory with general relativity. It is well known that Heisenberg suggested a unified quantum field theory of a fundamental spinor field describing all matter fields in their interactions\cite{hei1,hei2}. In his theory the masses and interactions of particles are a consequence of a self-interaction term of the elementary spinor field. The fact that manifolds with no-Euclidean geometry can help uncover new features of quantum matter makes it desirable to create manifolds of controllable shape and to develop the capability to add in synthetic gauge fields\cite{hohuang}. In the last decade new physical possibilities, beyond the standard Dirac, Majorana and Weyl spinors, have been introduced and studied in a Minkowsky 4D spacetime\cite{hoff}. In particular, symmetric curvature spinors with $2s$ indices that appear in twistors generalized Weyl curvature spinor ($s=2\hbar$) and were shown later to play an important role in the formulation of higher spin gauge theories extending Einstein\cite{vasiliev} and Weyl\cite{fradkin} gravity theories.

The pure spinor formalism was introduced by Berkovits\cite{Berk} as an alternative to the Green-Schwarz and BRST formalisms. Its main advantage are manifest space time supersymmetry and the fact that it can be quantised in the conformal gauge. This formalism was used in a wide range of applications\cite{Berk1,Berk2, Berk3, Chan, Vallilo, Chan1}. A natural question to ask is how to generalize the non-minimal spinor string to curved backgrounds, but the more important question is how to construct a string non-perturbative formalism on a curved background which can be independent of the gauge. An interesting suggestion was made in a curved spacetimes related to the Dirac equation by Saha and Shikin some years ago\cite{saha,saha1}, but without take into account self-interactions.

In this paper we construct a pure geometric spinor field theory on an arbitrary curved background, which is considered a Riemannian manifold. In the theory the spinor field $\hat{\Psi}^{\alpha}$ is responsible for the displacement of the extended Weylian manifold\cite{weyl} with respect to the Riemannian background. In our approach, the covariant derivative of the metric tensor in the Riemannian background manifold is null (we denote with a $\nabla$, the Riemannian-covariant derivative and with a $\Delta$ the Riemannian variation of some arbitrary tensor): $\Delta g_{\alpha\beta}=\nabla_{\gamma} g_{\alpha\beta} \,dx^{\gamma}=0$. However, the Weylian covariant derivative on the extended Weylian manifold (we denote the covariant derivative on the extended Weylian manifold with a $ \|  $ ), is nonzero: $ g^{\alpha\beta}_{\quad \|\gamma} \neq 0$. Furthermore, we shall consider the coupling of the spinor fields with the background and their self-interactions in a generic manner. The theory is worked in 8 dimensions, 4 of them related to the space-time coordinates ($x^{\mu}$), and the other 4 related to the inner space ($\phi^{\mu}$), described by compact coordinates. The former have the spin components as canonical momentums: ($s_{\mu}$).

\section{EINSTEIN-HILBERT ACTION AND QUANTUM STRUCTURE OF SPACE TIME}

If we deal with an orthogonal basis,
the curvature tensor will be written in terms of the connections:
$R^{\alpha}_{\,\,\,\beta\gamma\delta} = \Gamma^{\alpha}_{\,\,\,\beta\delta,\gamma} -  \Gamma^{\alpha}_{\,\,\,\beta\gamma,\delta}
+ \Gamma^{\epsilon}_{\,\,\,\beta\delta} \Gamma^{\alpha}_{\,\,\,\epsilon\gamma} - \Gamma^{\epsilon}_{\,\,\,\beta\gamma}
\Gamma^{\alpha}_{\,\,\,\epsilon\delta}$). The Einstein-Hilbert (EH) action for an arbitrary matter lagrangian density ${\cal L}$
\begin{equation}
{\cal I} = \int d^4 x \sqrt{-g} \left[ \frac{R}{2\kappa}+ {\cal L} \right],
\end{equation}
after variation, is given by
\begin{equation}\label{delta}
\delta {\cal I} = \int d^4 x \sqrt{-g} \left[ \delta g^{\alpha\beta} \left( G_{\alpha\beta} + \kappa T_{\alpha\beta}\right)
+ g^{\alpha\beta} \delta R_{\alpha\beta} \right],
\end{equation}
where $\kappa = 8 \pi G$, $G$ is the gravitational constant and $g^{\alpha\beta} \delta R_{\alpha\beta} =\delta\Theta$, such that
$\delta\Theta(x^{\alpha})$ is an arbitrary scalar field, and $T_{\alpha\beta}$ is the energy-momentum tensor defined by
\begin{equation}
T_{\alpha\beta}= 2 \frac{\delta{\cal L}}{\delta g^{\alpha\beta}} - g_{\alpha\beta} {\cal L}.
\end{equation}
When the flux $\delta\Theta(x^{\alpha})$ that cross the Gaussian-like hypersurface defined on an arbitrary region of the spacetime,
is zero, the resulting equations that minimize the EH action, are the background Einstein equations: $G_{\alpha\beta} + \kappa\, T_{\alpha\beta}=0$. However, when this flux is nonzero,
one obtains in the last term of the eq. (\ref{delta}). This flux becomes zero when there are no sources within this hypersurface. Hence, in order to make $\delta {\cal I}=0$ in Equation (\ref{delta}), we must consider the condition: $
G_{\alpha\beta} + \kappa T_{\alpha\beta} = \Lambda\,
g_{\alpha\beta}$, where $\Lambda$ is the cosmological constant. On the other hand, we can make the transformation
\begin{equation}\label{ein}
\bar{G}_{\alpha\beta} = {G}_{\alpha\beta} - \Lambda\, g_{\alpha\beta},
\end{equation}
where the scalar field $\delta\Theta$ complies $\Box \delta\Theta =0$\cite{rb}, and the transformed Einstein equations with the equation of motion for the transformed gravitational waves, hold
\begin{equation}
\bar{G}_{\alpha\beta} = - \kappa\, {T}_{\alpha\beta}. \label{e1} \\
\end{equation}
The Equation (\ref{e1}) give us the Einstein equations with cosmological
constant included. Notice that the scalar field $\delta\Theta(x^{\alpha})$ appears
as a scalar flux of some 4-vector with components $\delta W^{\alpha}$:
\begin{equation}\label{uchi}
\left[\delta W^{\alpha}\right]_{||\alpha} = \delta\Theta(x^{\alpha}),
\end{equation}
through the closed hypersurface $\partial{\cal M}$, which is situated in any region of space-time. Here,
$\delta W^{\alpha}=\delta\Gamma^{\epsilon}_{\beta\epsilon} g^{\beta\alpha}-\delta \Gamma^{\alpha}_{\beta\gamma} g^{\beta\gamma}$\footnote{We define the covariant derivative of some vector field $\Upsilon^{\beta}$: $\left[\Upsilon^{\beta}\right]_{||\alpha}$
\begin{equation}
\left[{\Upsilon}^{\beta}\right]_{||\alpha} = \nabla_{\alpha}{\Upsilon}^{\beta} + \left( \hat{\Psi}^{\beta} {\Upsilon}_{\alpha} -{\Upsilon}^{\beta}\hat{\Psi}_{\alpha}\right)+\xi^2 \,{\Upsilon}^{\beta} \hat{\Psi}_{\alpha},
\end{equation}
where $\xi$ is the self-interaction constant, $\nabla_{\alpha}\Upsilon^{\beta}$ is the covariant derivative on the Riemann manifold and $\delta\Gamma^{\beta}_{\epsilon\alpha}$ is the displacement of the manifold with respect to the Riemann one.}. In this work we shall introduce a extended manifold to describe quantum geometric spinor fields
$\hat{\Psi}^{\alpha}$, where the connections are
\begin{equation}\label{ga}
\hat{\Gamma}^{\alpha}_{\beta\gamma} = \left\{ \begin{array}{cc}  \alpha \, \\ \beta \, \gamma  \end{array} \right\}+ \hat{\Psi}^{\alpha}\,g_{\beta\gamma}.
\end{equation}
Here
\begin{equation}\label{uch}
\hat{\delta{\Gamma}}^{\alpha}_{\beta\gamma}=\hat{\Psi}^{\alpha}\,g_{\beta\gamma},
\end{equation}
describes the quantum displacement of the extended Weylian manifold with respect to the classical Riemannian background, which is described by the Levi-Civita symbols in (\ref{ga}), and the variation of the Ricci tensor is
\begin{equation}
\hat{\delta{R}}_{\beta\gamma} = \left(\hat{\delta\Gamma}^{\alpha}_{\beta\alpha} \right)_{\| \gamma} - \left(\hat{\delta\Gamma}^{\alpha}_{\beta\gamma} \right)_{\| \alpha},
\end{equation}
where $\hat{\delta\Gamma}^{\alpha}_{\beta\alpha}=\hat{\Psi}^{\alpha}\,g_{\beta\gamma}$. In the following, we will define the operator algebra of space-time in such a way that we can recover the metric on the manifold from it.

\subsection{Quantum structure of space-time}

In order to describe the quantum structure of space time we consider a the variation $\delta\hat{X}^{\mu}$ of the quantum operator $\hat{X}^{\mu}$, which describes the quantum space time coordinates, the variation $\delta\hat{\Phi}^{\mu}$ of the quantum operator $\hat{\Phi}^{\mu}$, that describes the quantum inner space. These operators can be applied to some background  quantum state, and describes a Fock space on an arbitrary Riemannian curved space time $\left|B\right>$, such that they comply with
\begin{equation}\label{dif}
\delta\hat{X}^{\mu}\left|B\right> = dx^{\mu}\left|B\right>, \qquad \delta\hat{\Phi}^{\mu}\left|B\right> = d\phi^{\mu}\left|B\right>.
\end{equation}
In order to describe the effective background space-time, we shall consider the line elements
\begin{equation}\label{line}
dx^2 \delta_{BB'}  = \left<B\right| \underrightarrow{\delta{X}} \overrightarrow{\delta{X}} \left| B'\right> , \qquad  d\phi^2 \delta_{BB'}=
\left<B\right| \underleftrightarrow{\delta\Phi} \overleftrightarrow{\delta\Phi}\left| B'\right>,
\end{equation}
where $\phi^{\alpha}$ are the four compact dimensions related to their canonical momentum components $s^{\alpha}$ that describe the spin. The variations and differentials of the operators $\hat{X}^{\mu}$ and $\hat{\Phi}^{\mu}$ on the extended Weylian manifold, are given respectively by
\begin{eqnarray}
\delta\hat{X}^{\mu}\left| B\right> &=& \left(\hat{X}^{\mu}\right)_{\|\alpha} dx^{\alpha}\left| B\right>, \qquad \delta\hat{\Phi}^{\mu} \left| B\right>= \left(\hat{\Phi}^{\mu}\right)_{\|\alpha} d\phi^{\alpha}\left| B\right>, \\
d\hat{X}^{\mu} \left| B\right>&=& \left(\hat{X}^{\mu}\right)_{,\alpha} dx^{\alpha}\left| B\right>, \qquad d\hat{\Phi}^{\mu} \left| B\right>= \left(\hat{\Phi}^{\mu}\right)_{,\alpha} d\phi^{\alpha}\left| B\right>,
\end{eqnarray}
with covariant derivatives that take into account the interaction of $\hat{X}^{\mu}$ with the geometrical spinor components $\hat{\Psi}^{\alpha}$\footnote{The covariant derivative of
$\hat{\Psi}^{\alpha}$ on an extended Weylian manifold with self-interaction, must be considered in the form
\begin{displaymath}
\left(\hat{\Psi}^{\alpha}\right)_{\|\beta} = \nabla_{\beta} \hat{\Psi}^{\alpha} + \hat{\Psi}^{\alpha} g_{\gamma\beta} \hat{\Psi}^{\gamma} +
\left(\xi^2-1\right)\hat{\Psi}^{\alpha} \hat{\Psi}_{\beta}=\nabla_{\beta} \hat{\Psi}^{\alpha}+\xi^2\,\hat{\Psi}^{\alpha} \hat{\Psi}_{\beta},
\end{displaymath}
where the coupling $\xi$ takes into account some physical properties that acts on the inner space (for instance, the colour charge) of the spinor fields under consideration.}:
\begin{eqnarray}
\left(\hat{X}^{\mu}\right)_{\|\beta}\left| B\right> &=& \left[\nabla_{\beta} \hat{X}^{\mu} + \hat{\Psi}^{\mu}  \hat{X}_{\beta} - \hat{X}^{\mu} \hat{\Psi}_{\beta}\right]\left| B\right>, \\
\left(\hat{\Phi}^{\mu}\right)_{\|\beta}\left| B\right> &=& \left[\nabla_{\beta} \hat{\Phi}^{\mu} + \hat{\Psi}^{\mu}  \hat{\Phi}_{\beta} - \hat{\Phi}^{\mu} \hat{\Psi}_{\beta}\right]\left| B\right>.
\end{eqnarray}
The comma denotes the ordinary partial derivative with respect to ordinary exterior coordinates. Hence, the requisite for the equations (\ref{dif}) to be fulfilled, are
\begin{eqnarray}
 \left[ \hat{X}^{\mu} \hat\Psi_{\beta} -\hat{\Psi}^{\mu} \hat{X}_{\beta}\right]\left|B\right>  & = & \left\{ \begin{array}{cc}  \mu \, \\ \epsilon \, \beta  \end{array} \right\} \hat{X}^{\epsilon} \left|B\right>, \\
 \left[ \hat{X}_{\mu} \hat\Psi^{\alpha} -\delta^{\alpha}_{\mu} \hat{\Psi}^{\epsilon} \hat{X}_{\epsilon}\right]\left|B\right>
& = & g^{\alpha\beta} \left\{ \begin{array}{cc}  \epsilon \, \\ \beta \, \mu  \end{array} \right\} \hat{X}_{\epsilon} \left|B\right>, \\
\left[  \hat{\Phi}^{\mu} \hat\Psi_{\beta} -\hat{\Psi}^{\mu} \hat{\Phi}_{\beta} \right]\left|B\right>
& = & \left\{ \begin{array}{cc}  \mu \, \\ \epsilon \, \beta  \end{array} \right\} \hat{\Phi}^{\epsilon} \left|B\right>, \\
\left[\hat{\Phi}_{\mu} \hat\Psi^{\alpha} -\delta^{\alpha}_{\mu} \hat{\Psi}^{\epsilon} \hat{\Phi}_{\epsilon}\right]\left|B\right> & = & g^{\alpha\beta} \left\{ \begin{array}{cc}  \epsilon \, \\ \beta \, \mu  \end{array} \right\} \hat{\Phi}_{\epsilon} \left|B\right>.
\end{eqnarray}

\subsection{Bi-vectorial structure of inner space}

We shall consider the squared of the $\hat{\delta\Phi}$-norm on the bi-vectorial space, and the squared $\hat{\delta X}$-norm on the vectorial space, are
\begin{eqnarray}
\underleftrightarrow{\delta\Phi} \overleftrightarrow{\delta\Phi} &\equiv& \left( \hat{\delta\Phi}_{\mu} \hat{\delta\Phi}_{\nu} \right) \left( \hat{\gamma}^{\mu} \hat{\gamma}^{\nu}\right), \\
\underrightarrow{\delta{X}} \overrightarrow{\delta{X}} & \equiv &  \hat{\delta{X}}_{\alpha} \hat{\delta{X}}^{\alpha}.
\end{eqnarray}
such that $\hat{\Phi}^{\alpha}=\phi\, \hat{\gamma}^{\alpha} $ and $\hat{X}^{\alpha}=x\, \hat{\gamma}^{\alpha} $ are respectively the components of the inner and coordinate spaces. Furthermore,  $\hat{\gamma}_{\mu}$ are the ($4\times 4$) Dirac matrices that generate the vectorial and bi-vectorial structure of the space time
\begin{eqnarray}
\left< B\left| \hat{X}_{\mu} \hat{X}^{\mu}\right|B\right> &=& x^2\, \,\mathbb{I}_{4\times 4}, \\
\left< B\left| \left( \hat{\Phi}_{\mu} \hat{\Phi}_{\nu} \right) \left( \hat{\gamma}^{\mu} \hat{\gamma}^{\nu}\right)\right|B\right> &=&
\left< B\left|\frac{1}{4} \left\{ \hat{\Phi}_{\mu} ,\hat{\Phi}_{\nu} \right\} \left\{ \hat{\gamma}^{\mu} ,\hat{\gamma}^{\nu}\right\} -\frac{1}{4}
\left[ \hat{\Phi}_{\mu}, \hat{\Phi}_{\nu} \right] \left[ \hat{\gamma}^{\mu}, \hat{\gamma}^{\nu}\right]\right|B\right> \nonumber \\
&=&\phi^2 \,\mathbb{I}_{4\times 4}. \nonumber
\end{eqnarray}
The $\hat\gamma^{\mu}$ matrices, comply with the Clifford algebra
\begin{equation}
\hat{\gamma}^{\mu} = \frac{\bf{I}}{3!} \left(\hat{\gamma}^{\mu}\right)^2\,\epsilon^{\mu}_{\,\,\alpha\beta\nu} \hat{\gamma}^{\alpha\beta}  \hat{\gamma}^{\nu} , \qquad \left\{\hat{\gamma}^{\mu}, \hat{\gamma}^{\nu}\right\} =
2 g^{\mu\nu} \,\mathbb{I}_{4\times 4}, \nonumber
\end{equation}
where ${\bf{I}}={\gamma}^{0}{\gamma}^{1}{\gamma}^{2}{\gamma}^{3}$, $\mathbb{I}_{4\times 4}$ is the identity matrix, $\hat{\gamma}^{\alpha\beta}=\frac{1}{2} \left[\hat{\gamma}^{\alpha}, \hat{\gamma}^{\beta}\right]$. In this paper we shall consider the following basis on a Minkowsky spacetime (in cartesian coordinates):
 $\left\{ \gamma^a,\gamma^b\right\} = 2 \eta^{ab} \mathbb{I}_{4\times 4}$
\begin{eqnarray}
&& \gamma^0= \,\left(\begin{array}{ll}  \mathbb{I} & 0 \\
0 &  -\mathbb{I} \ \end{array} \right),\qquad
\gamma^1=  \left(\begin{array}{ll} 0 &  -\sigma^1 \\
\sigma^1 & 0  \end{array} \right),  \nonumber \\
&& \gamma^2= \left(\begin{array}{ll} 0 &  -\sigma^2 \\
\sigma^2 & 0  \end{array} \right),  \qquad \gamma^3= \left(\begin{array}{ll} 0 &  -\sigma^3 \\
\sigma^3 & 0  \end{array} \right),\nonumber
\end{eqnarray}
such that the Pauli matrices are
\begin{eqnarray}
&& \sigma^1 = \left(\begin{array}{ll} 0 & 1 \\
1  & 0  \end{array} \right), \quad \sigma^2 = \left(\begin{array}{ll} 0 & -i \\
i  & 0  \end{array} \right), \quad \sigma^3 = \left(\begin{array}{ll} 1 & 0 \\
0  & -1  \end{array} \right). \nonumber
\end{eqnarray}

\section{SPINOR FIELD}

The  expression (\ref{uch}), with the connection (\ref{ga}), implies that
\begin{equation}\label{ps}
\frac{\hat{\delta W}^{\alpha}}{\delta l} = 3 \hat{\Psi}^{\alpha},
\end{equation}
$l$-being the Weylian 4-length.
The self-interacting effects do not necessary preserve the Riemannian flux of matter fields along the Gaussian hypersurface, so that $ \left(\hat{\delta W}^{\alpha}\right)_{\|\alpha}=\nabla_{\alpha} \hat{\delta W}^{\alpha}+\xi^2 \, \hat{\delta W}^{\alpha} \hat{\Psi}_{\alpha}\neq\nabla_{\alpha} \hat{\delta W}^{\alpha}$. Notice that when the coupling constant is zero: $\xi=0$, the Riemannian flux on the extended Weylian manifold is equal to the flux on the Riemannian one. The flux equation can be rewritten using (\ref{ps}), so that
\begin{equation}\label{lig}
\nabla_{\alpha} \hat{\Psi}^{\alpha}+\xi^2 \, \hat{\Psi}^{\alpha} \hat{\Psi}_{\alpha}=\frac{1}{3}\frac{\hat{\delta\Theta}}{\delta l}.
\end{equation}
However, when we describe matter fields, the coupling $\xi$ is nonzero. In general, $\xi$ depends on the theory under study (i.e. on the group representation of the spinor fields), and can be proportional to some physical property of the field (mass, charge, etc). Spinors with $\xi=0$, do not describe matter fields, but geometric fields.

In this framework, we can define respectively the slash and vector quantum fields $\slashed{\Psi}     = \hat{\Psi}_{\alpha} \hat{\gamma}^{\alpha}$, $\overleftrightarrow{\Psi} = \Psi_{\alpha} \hat{\gamma}^{\alpha}$. The $4$-vector components are $\hat\Psi_{\alpha}=\frac{\hat{\delta\Theta}}{\hat{\delta{\Phi}}^{\alpha}}$, where the flux of $\hat{\Psi}^{\alpha}$-field through the Gaussian hypersurface in eq. (\ref{lig}): $\hat{\Theta}\left( x^{\beta}|\phi^{\nu}\right)$, can be represented according to (\ref{line}), as a Fourier expansion in the momentum-space (the asterisk denotes the complex conjugate):
\begin{eqnarray}
\hat{\Theta}\left(x^{\beta}|\phi^{\nu}\right) &=& \frac{1}{(2\pi)^4} \int d^4k \int d^4 s   \nonumber \\
&\times & \left[ A_{s,k}\, \Theta_{k,s}(x^{\beta}) e^{\frac{i}{\hbar} \underleftrightarrow{S} \overleftrightarrow{\Phi}}
+ B^{\dagger}_{k,s} \, \Theta^*_{k,s}(x^{\beta}) e^{-\frac{i}{\hbar} \underleftrightarrow{S} \overleftrightarrow{\Phi}}\right]. \nonumber
\end{eqnarray}
For massless spinors it is fulfilled $\Box\Theta_{k,s}(x^{\beta})=0$, and we can define the spinor components $\hat{\Psi}_{\alpha}\left(x^{\beta}|\phi^{\nu}\right)$
\begin{eqnarray}
 \hat{\Psi}_{\alpha}&=& \frac{i}{\hbar (2\pi)^4} \int d^4k \int d^4s \frac{\delta \left(\underleftrightarrow{S} \overleftrightarrow{\Phi}\right)}{\hat{\delta\Phi}^{\alpha}} \left[ A_{s,k}\, \Theta_{k,s} e^{\frac{i}{\hbar} \underleftrightarrow{S} \overleftrightarrow{\Phi}} \right.\nonumber \\
&-&\left. B^{\dagger}_{k,s} \, \Theta^*_{k,s} e^{-\frac{i}{\hbar} \underleftrightarrow{S} \overleftrightarrow{\Phi}}\right], \nonumber
\end{eqnarray}
where
\begin{equation}
\frac{\delta }{\hat{\delta\Phi}^{\alpha}}\left(\underleftrightarrow{S} \overleftrightarrow{\Phi}\right) =  \left(2 g_{\alpha\beta}  \mathbb{I}_{4\times 4} - \hat{\gamma}_{\alpha} \hat{\gamma}_{\beta} \right) \hat{S}^{\beta} = 2 \hat{S}_{\alpha} - \hat{\gamma}_{\alpha} \,s,
\end{equation}
where $s\,\mathbb{I}_{4\times 4}= \frac{1}{4} \hat{S}_{\beta} \hat{\gamma}^{\beta}$. Additionally, the squared bi-vectorial $\hat{S}$-norm, is
\begin{equation}
\left\| \hat{S} \right\|^2 = \left<B\left|\underleftrightarrow{S} \overleftrightarrow{S} \right|B\right>= \frac{1}{4}\left<B\left|\left( \hat{S}_{\mu} \hat{S}_{\nu} \right) \left( \hat{\gamma}^{\mu} \hat{\gamma}^{\nu}\right)\right|B\right>=s^2
\mathbb{I}_{4\times 4},
\end{equation}
for $\hat{S}_{\mu} = s \,\bar{\gamma}_{\mu}$. In order to quantize the spin, we shall consider the universal invariant
\begin{equation}\label{invariant}
\underleftrightarrow{S} \overleftrightarrow{\Phi} = \frac{1}{4}\left( \hat{S}_{\mu} \hat{\Phi}_{\nu} \right) \left( \hat{\gamma}^{\mu} \hat{\gamma}^{\nu}\right)=s \phi \,
\mathbb{I}_{4\times 4} = (2\pi n \hbar) \,\mathbb{I}_{4\times 4},
\end{equation}
with $n$-integer. For this reason, gravitons (which have $s=2\hbar$), will be invariant under $\phi=n\,\pi$ rotations, vectorial bosons (with $s =\hbar$), will be invariant under $\phi= 2n\pi$) rotations, fermions with $s =\frac{1}{2} \hbar$ will be invariant under $\phi= 4 n \pi$ rotations, meanwhile fermions with $s =\frac{3}{2} \hbar$ are invariant under $\phi= \frac{4}{3} n \pi$ rotations. In the next section we will analize the fermionic bosonic field cases separately. Whether we work with one or the other will be determined by the commutation relations obeyed by the annihilation and creation operators in each case.

\section{FERMIONIC AND BOSONIC QUANTIZATION}

We shall consider the quantization of bosons and fermions, taking into account that they are geometric (Weylian), spinor fields. If we consider the case of fermions, we obtain, for $\hat{S}_{\mu}\hat{\gamma}^{\mu} =s\mathbb{I}_{4\times 4}$, the quantization expression
\begin{eqnarray}
&& \left< B\left| \left\{\hat{\Psi}_{\mu}({\bf x}, {\bf \phi}), \hat{\Psi}_{\nu}({\bf x}', {\bf \phi}') \right\}\right|B \right> \nonumber \\
&&= \frac{1}{2}\,\,\frac{s^2}{\hbar^2}\, \left\{\hat{\gamma}_{\mu}, \hat{\gamma}_{\nu} \right\}\,\mathbb{I}_{4\times 4} \,  \sqrt{\frac{\eta}{g}}  \,\,\delta^{(4)} \left({\bf x} - {\bf x}'\right) \,\delta^{(4)} \left({\bf \phi} - {\bf \phi}'\right),
\end{eqnarray}
where $\sqrt{\frac{\eta}{g}}$ is the squared root of the ratio between the determinant of the Minkowsky metric: $\eta_{\mu\nu}$ and the metric that describes the background: $g_{\mu\nu}$. This ratio describes the inverse of the relative volume of the background manifold. Furthermore, in the case of bosons, we obtain
\begin{eqnarray}
&& \left< B\left| \left[\hat{\Psi}_{\mu}({\bf x}, {\bf \phi}), \hat{\Psi}_{\nu}({\bf x}', {\bf \phi}') \right]\right|B \right> \nonumber \\
&& = \frac{s^2}{2 \hbar^2 } \left[\hat\gamma_{\mu} , \hat\gamma_{\nu}\right] \,  \sqrt{\frac{\eta}{g}} \,\,\delta^{(4)} \left({\bf x} - {\bf x}'\right) \,\delta^{(4)} \left({\bf \phi} - {\bf \phi}'\right).
\end{eqnarray}
This means that the expectation value for the expression (\ref{lig}) will be
\begin{equation}
\left<B\right| \nabla_{\alpha} \hat{\Psi}^{\alpha}\left|B\right> = \frac{1}{3} \left<B\right| \frac{\hat{\delta\Theta}}{\delta l}\left|B\right>,
\end{equation}
for bosons, and
\begin{equation}
\left<B\right| \nabla_{\alpha} \hat{\Psi}^{\alpha}\left|B\right> + \frac{4 \xi^2\,s^2}{\hbar^2}  \mathbb{I}_{4\times 4} \, \,  \sqrt{\frac{\eta}{g}} \,\,\delta^{(4)}\left(x-x'\right)\,\delta^{(4)} \left(\phi-\phi'\right)= \frac{1}{3} \left<B\right| \frac{\hat{\delta\Theta}}{\delta l}\left|B\right>,
\end{equation}
for fermions.

Additionally, we obtain the fields $\hat{\slashed{\Psi}}=\hat{\gamma}^{\mu} {\hat\Psi}_{\mu}$ and $\hat{\slashed{\Psi}}^{\dagger} = \left(\hat{\gamma}^{\mu} {\hat\Psi}_{\mu}\right)^{\dagger}$
\begin{eqnarray}
\hat{\slashed{\Psi}} & = &  -\frac{i \,\mathbb{I}_{4\times 4}}{(2^{3/4}\pi)^4 \hbar} \int d^4k \int d^4s \,s\, \left[ A_{s,k}\, \Theta_{k,s} e^{\frac{i}{\hbar} \underleftrightarrow{S} \overleftrightarrow{\Phi}} \right. \nonumber \\
&-& \left. B^{\dagger}_{k,s} \, \Theta^*_{k,s} e^{-\frac{i}{\hbar} \underleftrightarrow{S} \overleftrightarrow{\Phi}}\right], \\
\hat{\slashed{\Psi}}^{\dagger} & = & \frac{i \,\mathbb{I}_{4\times 4}}{(2^{3/4}\pi)^4\hbar} \int d^4k \int d^4s \,s\,\left[ A^{\dagger}_{s,k}\, \Theta^*_{k,s} e^{\frac{-i}{\hbar} \underleftrightarrow{S} \overleftrightarrow{\Phi}}\right. \nonumber \\
&-& \left. B_{k,s} \, \Theta_{k,s} e^{\frac{i}{\hbar} \underleftrightarrow{S} \overleftrightarrow{\Phi}}\right],
\end{eqnarray}
such that fermions and bosons obey the following algebraic rules
\begin{eqnarray}
&& \left< B\left| \left\{\hat{\slashed{\Psi}}({\bf x}, {\bf \phi}), \hat{\slashed{\Psi}}^{\dagger}({\bf x}', {\bf \phi}') \right\}\right|B \right> \nonumber \\
&&=  \frac{4 s^ 2 \,L^2_p}{\hbar^2} \left(|A_{k,s}|^2 + |B_{k,s}|^2\right)\,\,  \sqrt{\frac{\eta}{g}} \,\,\delta^{(4)} \left({\bf x} - {\bf x}'\right) \,\delta^{(4)} \left({\bf \phi} - {\bf \phi}'\right), \nonumber \\
&& \left< B\left| \left[\hat{\slashed{\Psi}}({\bf x}, {\bf \phi}), \hat{\slashed{\Psi}}^{\dagger}({\bf x}', {\bf \phi}') \right]\right|B \right> \nonumber \\
&& = \frac{4 s^2 \,L^2_p}{\hbar^2} \left(|A_{k,s}|^2 - |B_{k,s}|^2\right) \,\,  \sqrt{\frac{\eta}{g}} \,\,\delta^{(4)} \left({\bf x} - {\bf x}'\right) \,\delta^{(4)} \left({\bf \phi} - {\bf \phi}'\right),\nonumber
\end{eqnarray}
where, in order for the fields to be quantised, the following relationships must hold
\begin{eqnarray}
\frac{4 s^2 \,L^2_p}{\hbar^2 } \left(|A_{k,s}|^2 + |B_{k,s}|^2\right) &=&  \left(\frac{c^3 M^3_p}{\hbar}\right)^2,  \label{q1}\\
\frac{4 s^2 \,L^2_p}{\hbar^2} \left(|A_{k,s}|^2 - |B_{k,s}|^2\right) &=& 0, \,\pm \left(\frac{c^3 M^3_p}{\hbar}\right)^2. \label{q2}
\end{eqnarray}
The condition (\ref{q1}) is required for a successful quantization of spinor fermionic fields. The conditions (\ref{q2}) are required for scalar bosons (the first equality) and vector, or
tensor bosons (the second equality). On the other hand, in order for the expectation value of the energy to be positive: $\left< B\left| \hat{{\cal H}}\right|B \right> \geq 0$, we must choose the negative signature in the second equality of (\ref{q2}).

\subsection{RELATIVISTIC SPINOR DYNAMICS}

So far we've been working on the definitions of the algebra and geometry needed to incorporate the spinor field to the theory. The next is to determine the dynamics for such a field. In this approach, we consider the field as part of the Weyl connection, so we shall recover its dynamics from equations involving geometric tensors, such as the Einstein tensor of the Weyl-like manifold.

Because we need to describe a relativistic theory of spinor fields on an arbitrary curved space-time, we shall calculate the covariant derivative of the metric tensor, on the extended
Weyl manifold, as
\begin{equation}
\hat{g}_{\beta\alpha\|\gamma} =\nabla_{\gamma} g_{\beta\alpha} - \left(g_{\beta\gamma} \hat{\Psi}_{\alpha} + \hat{\Psi}_{\beta} g_{\alpha\gamma} \right) + \left(1-\xi^2\right)\left\{ \hat{\Psi}_{\gamma} , g_{\alpha\beta} \right\}, \nonumber
\end{equation}
where $\nabla_{\gamma} g_{\beta\alpha}=0$ is the covariant derivative on the Riemann manifold and we denote with  $\|$ the covariant derivative on the extended Weylian manifold which includes the massive spinor fields with mass $\xi$. Of course, in the cases in which the fields are massless bosons we must consider $\xi=0$. The terms between the parenthesis are due to the Weyl contributions and the last term is exclusively due to the self interaction of the spinor $\hat{\Psi}_{\nu}$. Se shall define
\begin{equation}
\hat{g}_{\beta\alpha\|\gamma} \,\hat{\delta X}^{\gamma} \left| B\right> = \delta g_{\beta\alpha} \left| B\right>,
\end{equation}
as the variation of the tensor metric due to the quantum disiplacement of the extended Weylian manifold with respect to the classical Riemannian background.

We are interested in obtaining the relativistic dynamics of the spinor fields $\hat{\Psi}_{\alpha}$ on the extended Weylian manifold. To do it, we must notice that the variation of the extended Ricci tensor $\hat{\delta{R}}^{\alpha}_{\beta\gamma\alpha}=\hat{\delta{R}}_{\beta\gamma}$, is
\begin{equation}
\hat{\delta{R}}_{\beta\gamma} = \left(\hat{\delta\Gamma}^{\alpha}_{\beta\alpha} \right)_{\| \gamma} - \left(\hat{\delta\Gamma}^{\alpha}_{\beta\gamma} \right)_{\| \alpha}, \nonumber
\end{equation}
such that (we denote with a {\em hat}, fields that are operators), we
obtain the following result for the quantum variation of the extended Ricci tensor with respect to the Riemannian one:
\begin{eqnarray}
\hat{\delta{R}}_{\beta\gamma} &=& \nabla_{\gamma} \hat{\Psi}_{\beta} - 3 \left(1-\frac{\xi^2}{3}\right) g_{\beta\gamma} \left(\hat{\Psi}^{\nu} \hat{\Psi}_{\nu}\right) \nonumber \\
&-&g_{\beta\gamma} \left(\nabla_{\nu} \hat{\Psi}^{\nu}\right) +\left(1-\frac{\xi^2}{3}\right)\hat{\Psi}_{\beta}\hat{\Psi}_{\gamma}.
\end{eqnarray}
This tensor field have symmetric $\hat{U}_{\beta\gamma}$, and antisymmetric $\hat{V}_{\beta\gamma}$, parts
\begin{eqnarray}
\hat{U}_{\beta\gamma}&=&\frac{1}{2} \left( \nabla_{\beta} \hat{\Psi}_{\gamma}+\nabla_{\gamma} \hat{\Psi}_{\beta}\right)  - g_{\beta\gamma} \left(\nabla_{\nu} \hat{\Psi}^{\nu}\right) \nonumber \\
&-& 3\left(1-\frac{\xi^2}{3}\right) g_{\beta\gamma} \left(\hat{\Psi}_{\nu} \hat{\Psi}^{\nu}\right)  + 3 \left(1-\frac{\xi^2}{3}\right) \left\{\hat{\Psi}_{\beta},\hat{\Psi}_{\gamma}\right\}, \nonumber \\
\hat{V}_{\beta\gamma} &=& -\frac{1}{2} \left( \nabla_{\beta} \hat{\Psi}_{\gamma}-\nabla_{\gamma} \hat{\Psi}_{\beta}\right)+ \frac{3}{2}\left(1-\frac{\xi^2}{3}\right) \left[\hat{\Psi}_{\beta}, \hat{\Psi}_{\gamma}\right] .\nonumber
\end{eqnarray}
It is easy to show that $\hat{\delta{R}}^{\alpha}_{\beta\alpha\gamma}=-\hat{\delta{R}}_{\beta\gamma}$, so that this tensor gives us redundant information.

On the other hand, the purely antisymmetric tensor
$\hat{\delta{R}}^{\alpha}_{\alpha\beta\gamma}\equiv \hat{\Sigma}_{\beta\gamma}$, is\footnote{Notice that (\ref{sigma}) can describe the gluon field strength tensor related to the gluon field components: $\hat{\Psi}^{\mu} = \frac{\lambda^n}{2}\, \hat{A}^{\mu}_n $, such that  $\lambda^n $ are the eight ($3\times 3$) Gell-Mann matrices in the $SU(3)$ group representation and $\xi^2=-(1+i\,g_s)$, such that $g_s=\sqrt{4\pi \alpha_s}$, $\alpha_s$ being the coupling constant of the strong force. The experimental value is $\alpha_s\simeq 0.1182$\cite{pdb}, that corresponds to $\xi=\pm(0.5369-1.135\,i)$. If we consider that $\hat{\Sigma}^{\beta\gamma}$ to be conserved on the extended Weylian manifold, we obtain the solution for a free gluon field: $\left(\hat{\Sigma}^{\beta\gamma}\right)_{\|\gamma}=0$, which can written in terms of Riemannian covariant derivatives
\begin{equation}\label{mov}
\nabla_{\gamma} \hat{\Sigma}^{\beta\gamma} + \hat{\Psi}_{\gamma} \hat{\Sigma}^{\beta\gamma} +\left(1-\xi^2\right) \left[\hat{\Psi}_{\gamma}, \hat{\Sigma}^{\beta\gamma}\right]=0.
\end{equation}
Furthermore, because in this example, we are considering a confined free gluon field, we must set $\hat{\delta{\Theta}}=0$ in the constraint equation (\ref{lig}):
\begin{equation}
\nabla_{\alpha} \hat{\Psi}^{\alpha}+\xi^2 \, \hat{\Psi}^{\alpha} \hat{\Psi}_{\alpha}=0,
\end{equation}
which, jointly with the equation (\ref{mov}), provide us the solution of the problem. For free gluon fields, we obtain the expectation value on the background
\begin{equation}
\lambda^n \left< B\left|\nabla_{\alpha} \hat{A}^{\alpha}_{n}\right| B\right>=0,
\end{equation}
because $g_{\alpha\beta} \left< B\left|\left\{\hat{\Psi}^{\alpha}, \hat{\Psi}^{\beta}\right\}\right| B \right> =0$ for bosons.}

\begin{equation}\label{sigma}
\hat{\Sigma}_{\beta\gamma} = \left( \nabla_{\beta} \hat{\Psi}_{\gamma}-\nabla_{\gamma} \hat{\Psi}_{\beta}\right) -\left(1+\xi^2\right) \left[ \hat{\Psi}_{\beta}, \hat{\Psi}_{\gamma} \right].
\end{equation}
It is possible to obtain the varied Einstein tensor on the extended Weylian manifold: $\hat{\delta{G}}_{\beta\gamma}= \hat{U}_{\beta\gamma} - \frac{1}{2} g_{\beta\gamma} \hat{U}$, where $\hat{U}=g^{\alpha\beta} \hat{U}_{\alpha\beta}$:
\begin{eqnarray}
\hat{\delta{G}}_{\beta\gamma}&=&\frac{1}{2} \left( \nabla_{\beta} \hat{\Psi}_{\gamma}+\nabla_{\gamma} \hat{\Psi}_{\beta}\right)+\frac{1}{2} g_{\beta\gamma}\left[ \left(1-\frac{\xi^2}{3}\right)\left(\hat{\Psi}^{\alpha}\hat{\Psi}_{\alpha}\right)\right. \nonumber \\
&+& \left. \left(\nabla_{\nu} \hat{\Psi}^{\nu}\right)\right]
+\frac{1}{2}  \left(1-\frac{\xi^2}{3}\right) \left\{\hat{\Psi}_{\beta}, \hat{\Psi}_{\gamma}\right\} . \label{gg}
\end{eqnarray}
Therefore, using the fact that [see the eq. (\ref{ein})] $\delta G_{\alpha\beta} = -g_{\alpha\beta}\, \Lambda$, we obtain that
\begin{equation}
\hat{\Lambda} = - \frac{3}{4} \left[ \nabla_{\alpha} \hat{\Psi}^{\alpha} + \left(1- \frac{\xi^2}{3}\right) \hat{\Psi}^{\alpha} \hat{\Psi}_{\alpha}\right],
\end{equation}
is the cosmological constant. As was demonstrated in a previous work\cite{rb}, $\Lambda$ is an invariant of Riemann, but not of Weyl, so that we must promote it to an quantum operator: $\hat{\Lambda}$, which has the Riemannian background expectation value
\begin{equation}
\left< B\left|\hat{\Lambda}  \right| B \right> = -\frac{3}{4} \left[ \left< B\left| \nabla_{\alpha} \hat{\Psi}^{\alpha} \right| B \right> + \frac{1}{2} g^{\alpha\beta}\left(1- \frac{\xi^2}{3} \right) \left< B\left| \left\{\hat{\Psi}_{\alpha}, \hat{\Psi}_{\beta} \right\}\right| B \right> \right].
\end{equation}
The last term indicates the fermionic contribution to the cosmological constant. The tensors $\hat{\Sigma}_{\beta\gamma}$ and $\hat{V}_{\beta\gamma}$ are both antisymmetric. In order to describe independently, massless spinor fields and matter spinor fields, we can make a linear combinations to define the tensors
$\hat{\cal{N}}_{\beta\gamma} = {1\over 2} \hat{V}_{\beta\gamma} - {1\over 4}\hat{{\Sigma}}_{\beta\gamma} $ and $\hat{\cal{M}}_{\beta\gamma} = {1\over 2} \hat{V}_{\beta\gamma} + {1\over 4} \hat{{\Sigma}}_{\beta\gamma} $
\begin{eqnarray}
\hat{\cal{N}}_{\beta\gamma} & = & -\frac{1}{2} \left( \nabla_{\beta} \hat{\Psi}_{\gamma}-\nabla_{\gamma} \hat{\Psi}_{\beta}\right) + \left[ \hat{\Psi}_{\beta}, \hat{\Psi}_{\gamma}\right], \label{n1}\\
\hat{\cal{M}}_{\beta\gamma} & = & \frac{1}{2} \left(1-\xi^2\right) \left[ \hat{\Psi}_{\beta}, \hat{\Psi}_{\gamma}\right], \label{n2}
\end{eqnarray}
such that the symmetric tensor $\hat{\delta{G}}_{\beta\gamma}$, with the antisymmetric ones $\hat{\cal{N}}_{\beta\gamma}$ and $\hat{\cal{M}}_{\beta\gamma}$, describe all the dynamics of possible spinor fields, which must be conserved on the extended Weylian manifold
\begin{equation}
\left(\hat{\delta{G}}^{\beta\gamma}\right)_{\|\gamma} =0, \quad \left(\hat{\cal{N}}^{\beta\gamma}\right)_{\|\gamma}=0, \quad \left(\hat{\cal{M}}^{\beta\gamma}\right)_{\|\gamma}=0.
\end{equation}
This means that they must comply with the equations
\begin{eqnarray}
\nabla_{\beta} \left(\hat{\delta{G}}^{\alpha\beta}\right) &+& \left(\xi^2-1\right)\left\{ \hat{\Psi}_{\mu}, \hat{\delta{G}}^{\alpha\mu}\right\} =- 3  \hat{\Psi}_{\mu} \hat{\delta{G}}^{\alpha\mu}, \label{a} \\
\nabla_{\beta} \left(\hat{\cal{M}}^{\alpha\beta}\right) &+& \xi^2 \left[ \hat{\Psi}_{\mu}, \hat{\cal{M}}^{\alpha\mu}\right]
= -    \hat{\cal{M}}^{\alpha\mu}\hat{\Psi}_{\mu}, \label{b} \\
\nabla_{\beta} \left(\hat{\cal{N}}^{\alpha\beta}\right) &=&-     \hat{\cal{N}}^{\alpha\mu}\hat{\Psi}_{\mu}. \label{c}
\end{eqnarray}
The anti-commutator in (\ref{a}) is due to the symmetry of the tensor $\hat{\delta{G}}^{\alpha\beta}$. These
algebraic terms are of an self-interactive nature, on the extended Weylian manifold. The anti-symmetric bi-vectorial field $\hat{\cal{N}}^{\alpha\beta}$, describes the graviton fields and their dynamics is determined by the equation (\ref{c}). The terms on the right side of (\ref{a}-\ref{c}) can be considered as effective sources on the background Riemannian manifold.

Explicitly written, the equations (\ref{b}) and (\ref{c}), describe respectively the dynamics for charged and neutral vector bosons
\begin{eqnarray}
\left(\nabla_{\gamma} \hat{\Psi}^{\beta}\right)\hat{\Psi}^{\gamma} -\left(\nabla_{\gamma} \hat{\Psi}^{\gamma}\right)\hat{\Psi}^{\beta} &-& \xi^2 \left[ \hat{\Psi}_{\gamma}, \left[
\hat{\Psi}^{\beta},\hat{\Psi}^{\gamma}\right]\right] \nonumber \\
+  \left[ \hat{\Psi}^{\beta},\hat{\Psi}^{\gamma}\right] \hat{\Psi}_{\gamma} &+& \hat{\Psi}^{\beta} \left(
\nabla_{\gamma} \hat{\Psi}^{\gamma} \right)
= \hat{\Psi}^{\gamma} \left(\nabla_{\gamma} \hat{\Psi}^{\beta}\right), \\
\Box {\hat{\Psi}}^{\alpha}-\nabla_{\beta} \left(\nabla^{\alpha} \hat{\Psi}^{\beta} \right) +  &2& \left( \nabla_{\beta} \hat{\Psi}^{\alpha}\right) \hat{\Psi}^{\beta} - 2 \left(
\nabla_{\beta} \hat{\Psi}^{\beta} \right) \hat{\Psi}^{\alpha} \nonumber \\
- \left(\nabla^{\alpha}\hat{\Psi}^{\gamma}\right) \hat{\Psi}_{\gamma} + &2& \hat{\Psi}^{\alpha} \left(
\nabla_{\gamma} \hat{\Psi}^{\gamma} \right) + \left(\nabla^{\gamma}\hat{\Psi}^{\alpha}\right)\hat{\Psi}_{\gamma}   \nonumber \\
-2 \hat{\Psi}^{\gamma} \left(\nabla_{\gamma} \hat{\Psi}^{\alpha}\right) &=&  2 \left[ \hat{\Psi}^{\mu},
\hat{\Psi}^{\alpha} \right] \hat{\Psi}_{\mu} ,  \label{a2}
\end{eqnarray}

\section{AN EXAMPLE: CONFINEMENT OF GRAVITONS ON PLANCKIAN SCALES}

As an example, we can consider the interesting case where the system is asymptotically freedom on very small scales (with respect to the Planckian scales) and can be described by the
line element
\begin{equation}\label{mm}
dl^2= \pm\left[\left(1-\Lambda r^2\right) dt^2 - \frac{dr^2}{\left(1-\Lambda r^2\right)} - r^2 d\Omega^2\right]+ \left\| \hat{\delta \Phi} \right\|^2,
\end{equation}
such that $d\Omega^2 = d\theta^2 + \sin^2{\theta}\,d\varphi^2$. Here, $\Lambda=3 l^{-2}_p$, is the cosmological constant related to Planckian scales, and $l_p = 1.61 \times 10^{-35}\, m$ is the Planckian length. For $r< \Lambda^{-1/2}$ we must assume the positive signature in (\ref{mm}) and for $r> \Lambda^{-1/2}$ the negative signature. On Planckian scales it is pertinent
to consider gravitons, so that we shall consider $s=2\hbar$, and therefore: $\hat{S}_{\mu} = s\,\bar{\gamma}_{\mu}= 2 \hbar \bar{\gamma}_{\mu}$, such that the matrices on this metric are:
$\bar{\gamma}_{\mu}= e^a_{\mu}{\gamma}_{a}$, and  $e^a_{\mu}$ are the vielbein:
\begin{equation}
e^a_{\mu}= \left( \begin{array}{llll}  \sqrt{f(r)} & 0 & 0 & 0 \\
0 &  \frac{\sin{\theta}
\cos{\varphi}}{\sqrt{f(r)}} &
\frac{\sin\theta \sin\varphi}{\sqrt{f(r)}} &
\frac{\cos\theta}{\sqrt{f(r)}}
 \\
0 &  r\,\cos\theta \cos\varphi &
 r\,\cos\theta \sin\varphi & -
 r\,\sin\theta
 \\
0 & - r\,\sin\theta \sin\varphi &
 r\,\sin\theta
\cos\varphi & 0 \end{array} \right), \nonumber
\end{equation}
where $f(r)=\left(1-\Lambda r^2\right)$.  In order to study the localization of gravitons on the background Riemannian metric we shall consider the relativistic equation: $P_{\mu} P^{\mu} = m^2_0$, where $m_0=\xi \,M_p$ is the rest mass
of gravitons, and $M_p=1.2\times 10^{19}\,{\rm GeV}$ is the Planckian mass. For massless gravitons: $m_0=0$. To obtain a dynamical equation we must promote $P_{\mu}$ and $P^{\mu}$ to differential operators: $\hat{P}_{\mu} =i\,\hbar \nabla_{\mu}$, $\hat{P}^{\mu} =-i\,\hbar \nabla^{\mu}$, so that, if we apply these operators to the wave function $\Theta_{\omega,j,j_z}$, we obtain the differential equation
\begin{equation}
\Box  \Theta_{\omega,j,j_z}(x^{\beta})  = \left(\frac{m_0}{\hbar}\right)^2 \Theta_{\omega,j,j_z}(x^{\beta}) =0.\nonumber
\end{equation}
Now, because the elements $g_{\mu\nu}$ are does not depend on time, the energy will be a conserved quantity: $\Theta_{\omega,j,j_z}(x^{\beta})=\Theta^{(0)}_{\omega,j,j_z} (\vec{x})\, e^{\pm i \,\frac{E}{\hbar} t}$, so that
\begin{eqnarray}
\frac{\partial^2 \Theta_{\omega,j,j_z}(x^{\beta})}{\partial t^2}& = &\pm \frac{i\,E}{\hbar} \frac{\partial \Theta_{\omega,j,j_z}(x^{\beta})}{\partial t}\nonumber \\
&=& \left(\pm \frac{i\,E}{\hbar}\right)^2 \Theta_{\omega,j,j_z}(x^{\beta}) = - \frac{E^2}{\hbar^2} \Theta_{\omega,j,j_z}(x^{\beta}), \nonumber
\end{eqnarray}
so that with these requirements, we obtain the effective Schr\"odinger equation
\begin{equation}
g^{ij} \nabla_i \nabla_j \Theta_{\omega,j,j_z}(x^{\beta}) = \frac{E^2}{\hbar^2 f(r)} \Theta_{\omega,j,j_z}(x^{\beta}), \nonumber
\end{equation}
where $\omega=E/\hbar$.

If we use separation of variables: $\Theta_{\omega,j=l+s,j_z=m+s_z}(x^{\beta})= T_{\omega}(t) \,R_{\omega,j}(r)\,Y_{j,j_z}(\theta,\varphi)$, we obtain the solutions
\begin{eqnarray}
T_{\omega}(t) & = & e^{\pm i\omega t}, \nonumber \\
R_{\omega,j}(r) & = &  \left(r^2 \Lambda -1 \right) \nonumber \\
&\times& \left\{ D_1 \, r^{-(j+1)} \, {\cal{H}}\left[\left[A_1,B_1\right],[\frac{1}{2} -j], r^2 \Lambda \right] \right. \nonumber \\
&+& \left. D_2 \, r^{j}
{\cal{H}}\left[\left[A_2,B_2\right],[j+\frac{3}{2}], r^2 \Lambda \right]\right\} , \label{radial}  \\
Y_{j,j_z}(\theta,\varphi)& = & e^{i j_z \varphi} \, \sqrt{(2j+1)\frac{(j-j_z)!}{(j+j_z)!}}\,{\cal P}_{j,j_z}\left(\cos\theta\right), \nonumber
\end{eqnarray}
where ${\cal P}_{j,j_z}\left(\cos\theta\right)$ are the Legendre polynomials and ${\cal{H}}$ denotes the hypergeometric function with parameters
\begin{eqnarray}
A_1 & = &  \frac{-\left[i \omega + (j+1) \sqrt{\Lambda}\right]}{2\sqrt{\Lambda}},  \,\, A_2=\frac{-\left[i \omega -j \sqrt{\Lambda}\right]}{2\sqrt{\Lambda}}, \nonumber \\
B_1 & = &  \frac{-\left[i \omega + (j-2) \sqrt{\Lambda}\right]}{2\sqrt{\Lambda}},  \,\, B_2=\frac{-\left[i \omega -(j+3) \sqrt{\Lambda}\right]}{2\sqrt{\Lambda}}. \nonumber
\end{eqnarray}
In absence of angular moment (because the elements of the background metric tensor are isotropic), the only variable that can broke the spherical symmetry is the spin. Therefore, we must set $l=m=0$, so that $j=s$ and $j_z=s_z$. Hence, the spinor $ \hat{\Psi}_{\alpha}\left(t,r,\theta,\varphi|\phi^{\nu}\right)$ will be expanded as
\begin{eqnarray}
\hat{\Psi}_{\alpha}&=& \frac{-i}{2\hbar (2\pi)} \int d\omega \int d^4s \,\bar{\gamma}_{\alpha}\,s\, \left[ A_{s,\omega}\, \Theta_{\omega,s,s_z} e^{\frac{i}{\hbar} \underleftrightarrow{S} \overleftrightarrow{\Phi}}\right. \nonumber \\
& - & \left. B^{\dagger}_{\omega,s} \, \Theta^*_{\omega,s,s_z} e^{-\frac{i}{\hbar} \underleftrightarrow{S} \overleftrightarrow{\Phi}}\right],
\end{eqnarray}
where $\Theta_{\omega,s,s_z}(t,r,\theta,\varphi)= T_{\omega}(t) \,R_{\omega,s}(r) \,Y_{s,s_z}(\theta,\varphi)$. In the figure (\ref{f1}) is shown $R_{\omega,s=2\hbar}(r)$ inside of a Planckian radius. On the other hand, in the figures (\ref{f2}) and (\ref{f3}) are respectively plotted the real and imaginary parts of the spherical harmonic $Y_{s=2,s_z=2}(\theta,\varphi)$. Of course, all the distributions $\Theta_{\omega,s=2,s_z=\pm 2}(t,r,\theta,\varphi)$ oscillate with some frequency, which we are considered as $\omega=2.1436\times 10^{36}\quad{\rm sec}^{-1}$ in the plot.

In order for the spinor to be quantised we must require that the condition (\ref{q2}) to be fulfilled, so that
\begin{equation}
\frac{4 s^2 }{\hbar^2} \left(|A_{\omega,s}|^2 - |B_{\omega,s}|^2\right) =- \left(\frac{c^3 M^3_p}{\hbar\,L_p}\right)^2,
\end{equation}
where in the case of gravitons: $s=2\hbar$ and $s_z=\pm 2\hbar$. Therefore, the relationship between coefficients is
\begin{equation}
\left< B \left| A_{\omega,s}\, A^{\dagger}_{\omega,s}\right| B\right> = \left< B \left| B_{\omega,s} \,B^{\dagger}_{\omega,s}\right|B\right> - \left(\frac{c^3 M^3_p \,  }{4\,\hbar\,L_p}\right)^2.
\end{equation}
In order $R_{\omega,s}(r)=0$ at $r=0$, we must require that $D_1=0$ in (\ref{radial}). The antisymmetric tensor $\hat{\cal{N}}_{\alpha\beta}$ in (\ref{n1}), describes, for spin $s=2\hbar$, the graviton fields. The expectation value on the Riemannian background is nonzero for $\mu\neq \beta$:
\begin{eqnarray}
\left< B \left| \hat{\cal{N}}_{\beta\mu}\right|B\right>& = &  \frac{2}{L^2_p}\, \left[\bar\gamma_{\beta} , \bar\gamma_{\mu}\right] \, \delta^{(4)} \left({\bf x} - {\bf x}'\right) \,\delta^{(4)} \left({\bf \phi} - {\bf \phi}'\right), \nonumber
\end{eqnarray}
because $\left< B \left| \partial_{\beta} \hat{\Psi}_{\mu}-\partial_{\mu} \hat{\Psi}_{\beta}\right|B\right> =0$. The bi-vectorial field (graviton field) $\hat{\cal{N}}_{\beta\mu}$, must meet the dynamical equation (\ref{c}), for the extended Einstein tensor (\ref{gg}), with $\mu=\beta$.

\section{FINAL COMMENTS}

We have studied a non-perturbative theory of unification for quantised spinor fields on extended Weylian manifolds, taking into account the self-interactions of the spinors. The extended Weylian spacetime is represented by eight dimensions, the standard four exterior dimensions and other four inner dimensions (one by each exterior dimension). Each component of spin $\hat{S}_{\mu}$, is defined as the momentum corresponding to the inner dimension $\hat{\Phi}^{\mu}$, such that one can define an universal bi-vectorial invariant [see eq. (\ref{invariant})]: $\left<B\left|\underleftrightarrow{S} \overleftrightarrow{\Phi}\right|B\right> = (2\pi n \hbar) \,\mathbb{I}_{4\times 4}$. As an example we have examined the confinement of gravitons on a de Sitter background (static) metric, such that the cosmological constant assumes Planckian values. The exact results are surprising and the modes of the graviton field were summarized in (\ref{radial}).

The theory can be applied to many examples in the framework of quantum field theory, gravitation and cosmology. Many of these applications will be developed in further works.

\newpage
\begin{figure}[h]
\noindent
\includegraphics[width=.6\textwidth]{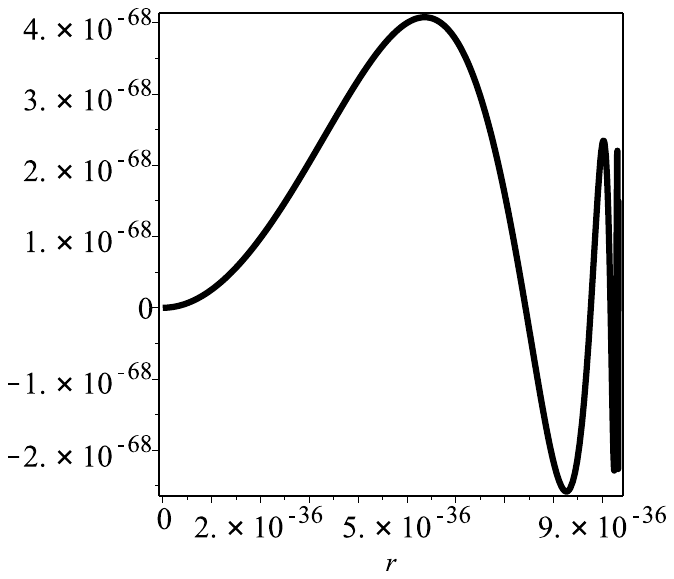}\vskip -8cm\caption{Plot of $R_{\omega,s}(r)$ for the graviton spinor with $s=2\hbar$ and $\omega=2.1436\times 10^{36}\quad{\rm sec}^{-1}$. Notice that $R_{\omega,s}(r=0)=0$, as one expects for a radial quadrupole distribution. The field is confined to the Planckian length $l_p=1.6\times 10^{-35}$ meters.}\label{f1}
\end{figure}
\vskip 2cm
\begin{figure}[h]
\noindent
\includegraphics[width=.7\textwidth]{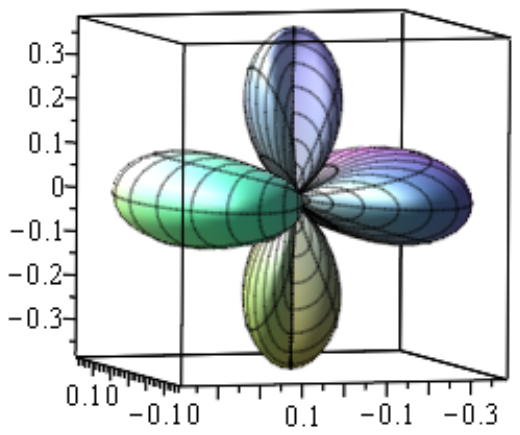}\vskip -9cm\caption{Plot of ${\rm{ Re}}\left[Y_{2,2}(\theta,\varphi)\right]$ for the graviton spinor.}\label{f2}
\end{figure}
\vskip 5cm
\begin{figure}[h]
\noindent
\includegraphics[width=.7\textwidth]{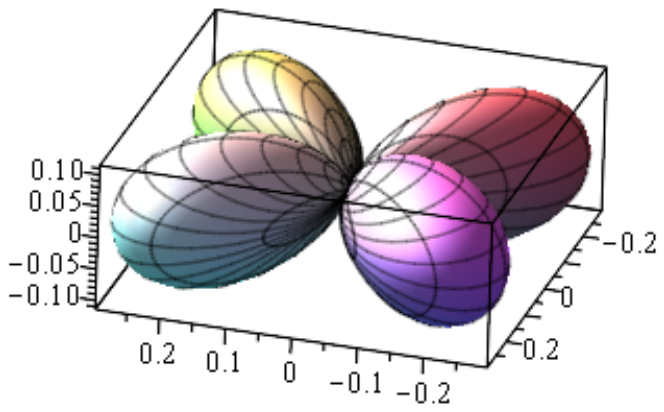}\vskip -9cm\caption{Plot of ${\rm{ Im}}\left[Y_{2,2}(\theta,\varphi)\right]$ for the graviton spinor.}\label{f3}
\end{figure}

\end{small}
\end{document}